\documentclass[12pt,letterpaper]{JHEP3}
\usepackage{cite}
\usepackage{epsfig}
\usepackage{amsfonts}

\title{Holographic Domains of Anti-de~Sitter Space}

\author{
Raphael Bousso \\
Institute for Theoretical Physics\\
University of California, Santa Barbara, CA 93106-4030, U.S.A.\\
E-mail: \email{bousso@itp.ucsb.edu}}
\author{
Lisa Randall \\
Department of Physics, Jefferson Laboratory\\
17 Oxford Street\\
Harvard University, Cambridge, MA 02138, U.S.A.\\
E-mail: \email{randall@physics.harvard.edu}}

\abstract{An AdS$_4$ brane embedded in AdS$_5$ exhibits the novel
feature that a four-dimensional graviton is localized near the brane,
but the majority of the infinite bulk away from the brane where the
warp factor diverges does not see four-dimensional gravity.  A naive
application of the holographic principle from the point of view of the
four-dimensional observer would lead to a paradox; a global
holographic mapping would require infinite entropy density. In this
paper, we show that this paradox is resolved by the proper covariant
formulation of the holographic principle. This is the first explicit
example of a time-independent metric for which the spacelike
formulation of the holographic principle is manifestly
inadequate. Further confirmation of the correctness of this approach
is that light-rays leaving the brane intersect at the location where
we expect four-dimensional gravity to no longer dominate. We also
present a simple method of locating CFT excitations dual to a particle
in the bulk.  We find that the holographic image on the brane moves
off to infinity precisely when the particle exits the brane's
holographic domain.  Our analysis yields an improved understanding of
the physics of the AdS$_4$/AdS$_5$ model.}

\preprint{\hepth{0112080} \\ NSF-ITP-01-167 \\ HUTP-01/A062}

\begin{document}

\section{Introduction}
\label{sec-intro}

The holographic principle~\cite{Tho93,Sus95} has frequently been
interpreted in a space-like sense.  The space-like formulation asserts
that the entropy in a spatial volume $V$ enclosed by a surface of area
$A$ cannot exceed $A/4$, in Planck units.  At each instant of time,
the complete information about physics in $V$ can be encoded on the
boundary $A$, at a density of no more than one bit per Planck area.
The space-like formulation is ambiguous as it involves an arbitrary
coordinate choice.  The area $A$ and the volume $V$ are fully
specified only after a particular equal-time slice has been picked.

More significantly, in strongly time-dependent backgrounds the
space-like holographic principle can actually be violated.
Counter-examples were first identified in cosmological
spacetimes~\cite{FisSus98}.  The entropy of sufficiently large
super-horizon regions will exceed their surface area.  More generally,
violations are easily found among systems in which gravitational
expansion or contraction dominates the dynamics~\cite{Bou99b}.  Inside
a black hole, for example, a star undergoing gravitational collapse
will have arbitrarily small surface area shortly before it is crushed
in a singularity.

Building on~\cite{FisSus98}, these difficulties were resolved by a
covariant reformulation~\cite{Bou99b,Bou99c,Bou99d}.  In the covariant
holographic principle, the entropy bound, $S \leq A/4$, is not
modified.  But instead of the entropy on a space-like hypersurface,
one considers the entropy on a null hypersurface, or ``light-sheet''.
Given an area $A$ (any open or closed codimension 2 spatial surface),
a light-sheet is generated by contracting light-rays emanating
orthogonally from $A$.  

No counter-example to the covariant entropy bound is known.  What
further distinguishes the light-sheet construction from previous
formulations of the holographic principle is its ability to relate
{\em any\/} area $A$ uniquely to adjacent bulk regions, even if $A$ is
not closed.  (This particular feature will be crucial to our
construction of holographic bulk regions associated with branes.)  A
better understanding of the underlying reason for the success of the
light-sheet formulation might well lead to new insights into the
connection between classical geometry and its statistical origin in a
quantum theory of gravity.

The covariant holographic principle is intriguing conceptually and has
proven essential in time-dependent backgrounds.  However, the
covariant formulation has not yet been brought to bear on holography's
most detailed manifestation, the AdS/CFT correspondence.  For certain
classes of asymptotically anti-de~Sitter space-times, the AdS/CFT
correspondence~\cite{Mal97,Wit98a,GubKle98} asserts that the physics
in the bulk of the space-time is fully described by a conformal field
theory (CFT) on the boundary.  The number of degrees of freedom of the
CFT is in accordance with the holographic bound~\cite{SusWit98}.

In the AdS/CFT correspondence, no contradictions to the space-like
formulation had been noticed until now.  At least from a global point
of view, this is not surprising.  Under the {\em space-like
condition\/} (the existence of a light-sheet that is {\em complete},
i.e., has no boundaries other than $A$), the covariant holographic
principle implies the space-like holographic principle as a special
case~\cite{Bou99b}.  The space-like condition is met by some%
\footnote{Note that any system can be surrounded by surfaces of
arbitrarily small area, by choosing a suitably wiggled equal-time slice.
(One can obtain such slices by piecing together sections of equal time
surfaces of observers moving nearly at the speed of light relative to
the system.)  This is a somewhat unnatural, but perfectly admissible
choice, and it obviously leads to a violation of the space-like
holographic principle (though not of the covariant version).  However,
the break-down of the space-like holographic principle exposed in this
paper is unrelated to this particular weakness.}
closed surfaces around any isolated, weakly gravitating matter system.
In a perturbative treatment of AdS, the entire space satisfies the
space-like condition~\cite{Bou99c}; a sphere near spatial infinity
possesses a complete light-sheet.%
\footnote{Space-like holography need not be challenged even when black
holes are allowed in the bulk.  By black hole
complementarity~\cite{SusTho93}, the CFT will be in one-to-one
correspondence with the portion of the bulk consisting of the black
hole exterior and the black hole horizons.  This part of the geometry
is not strongly time-dependent, and the space-like holographic
principle is expected to hold.}

Space-like holography has also been sufficient for discussions of the
Randall-Sundrum (RS) model~\cite{RS}.  This model describes a brane
whose metric is 3+1 dimensional Minkowski space.  The brane moves in a
portion of AdS$_5$, at fixed radial position in Poincar\'e
coordinates.  From the AdS/CFT correspondence, one expects that
objects in the bulk can alternatively be described by excitations of a
CFT on the brane~\cite{GidKat00a}.  The CFT must be coupled to the
four-dimensional graviton localized on the brane.  In
Sec.~\ref{sec-rs} we explain why space-like holography has not led to
contradictions in this model: It implies entropy bounds that are
actually weaker than those obtained from the covariant formulation.

However, from the vantage point of space-like holography, the case of
an AdS$_4$ brane embedded in AdS$_5$ (KR)~\cite{KarRan00} poses a
puzzle.  The KR model presents a significant new development that
broadens greatly the class of theories that might exhibit
localization.  It illustrates that the localization of gravity does
not depend on certain restrictive conditions satisfied by the RS
model.  These include the necessity for a finite volume bulk space and
a decreasing warp factor as an asymptotic boundary condition.  The
localization occurs due to local features of the geometry; regions far
from the brane are not relevant. This is discussed in detail
in~\cite{KarRan00}.

In the RS model, the graviton zero mode needs to be incorporated as an
explicit addition to the original CFT and to the original
five-dimensional space.  One remarkable property of the KR model is
that from the point of view of the boundary CFT, the four-dimensional
graviton is also a bound state~\cite{KarRan01b,Por01}.  Moreover, the
KR model displays features not previously seen in any four-dimensional
theory of gravity.  First, the graviton has a small mass, whose value
is set by the curvature scale of the AdS$_4$.

Second, in the KR model four-dimensional gravity dominates only over a
fraction of the volume.  The graviton appears as a light bound state
that only extends over a finite region of the bulk space.  This means
that observers localized to different regions of the space would see
the dimensionality of the space as different.

However, the space-like holographic principle will not reflect this
distinction. In fact, we show in Sec.~\ref{sec-ads} that space-like
holography would assign a bulk region of infinite entropy to a finite
area region on the brane, clearly a contradiction.

Some clue to the resolution of this paradox is that the boundary of
the bulk space is not only the AdS$_4$ brane, but also half of the
AdS$_5$ boundary.  One might expect different holographic regions to
be associated with the different portions of the boundary.  This is
particularly relevant as only part of the space even possesses
four-dimensional gravity, and the holographic correspondence would
need to distinguish such a region.

A further question that arises in the context of the KR model is the
interpretation of Newton's constant. In the RS model, $G_{\rm N}$ is
determined by the normalization of the zero mode, which is in turn
determined by the volume of the bulk space.  By contrast, the volume
of the bulk in the KR model is infinite and clearly does not
correspond to the graviton normalization.

In the main part of our paper, we show that the resolution of these
difficulties lies in the covariant formulation of the holographic
principle.  This is important in that it is the first time-independent
space for which this distinction is essential.  We will also see
evidence that the covariant treatment gives the correct description of
the physics of the AdS$_4$/AdS$_5$ system, increasing our confidence
both in the consistency of the KR model and in the covariant
formulation itself.

Furthermore, we find the four-dimensional region is essentially
equivalent to the holographic domain assigned to the brane. This is
also the region whose volume yields the proper graviton normalization
and therefore $G_{\rm N}$.

In Sec.~\ref{sec-ls}, we review the covariant entropy bound and propose
its application to branes.  The construction of light-sheets starting
at a given area is described in detail in Sec.~\ref{sec-ls1}.
Light-sheets must be terminated at caustics, when the generating
light-rays intersect.  In Sec.~\ref{sec-ls2}, we explain how to locate
caustics, and we show how this calculation simplifies under certain
conditions.  A brane is not an area but a time-like hypersurface; it
contains a whole sequence of areas, giving rise to a sequence of
light-sheets.  Sec.~\ref{sec-ls3} describes our adaption of the
covariant entropy bound to this case.  After resolving a slicing
ambiguity, we define the holographic domain, $\cal D$, of a given
brane as the bulk region covered by the sequence of light-sheets.
This is the bulk region whose physics we may expect to be dual to
holographic states on the brane.

In Sec.~\ref{sec-domains}, we apply the covariant formalism to an
AdS$_4$ brane in AdS$_5$.  We determine the caustics of light-sheets
that start on the brane, and we find that they occur at a finite
distance from the brane.  Thus we demonstrate that the holographic
domain associated with an AdS$_4$ brane does not include the entire
AdS$_5$ bulk, but extends only from the brane to another AdS$_4$
hypersurface near the minimum of the warp factor.  The remaining
portion of the bulk constitutes half of the AdS$_5$ space-time.  The
physics in this portion cannot be holographically represented on the
brane.

This raises an important question.  Consider a bulk particle moving
away from the brane.  When the particle crosses the minimum of the
warp factor, it leaves the holographic domain of the brane and should
no longer have a CFT dual on the brane.  But how can the CFT
excitation dual to the particle actually disappear from the brane?

In Sec.~\ref{sec-diamonds}, we develop an extremely simple
prescription relating the position of a bulk excitation to the locus
of the dual CFT excitation on a brane.  We use only causality,
employing the tool of causal diamonds.  Causal diamonds describe the
space-time region probed in an arbitrary experiment of given
duration~\cite{Bou00a}.  We find the beginning and end points of a
brane observer's experiment that is just barely long enough to detect
a given excitation in the bulk.  The conjectured equivalence of bulk
and boundary descriptions implies that the same experiment must also
barely detect the dual CFT excitation on the brane.  Hence the CFT
state has support on the boundary of the associated causal diamond.

In Sec.~\ref{sec-evanescence}, we apply the causal diamond method to
branes in AdS$_5$.  In Sec.~\ref{sec-rsc}, we consider the RS brane.
We verify that the CFT image predicted by causal diamonds agrees with
previous results obtained by more elaborate methods.  The image is a
shell, whose radius grows with the distance of a bulk excitation from
the brane.  In Sec.~\ref{sec-adsc}, we find the same qualitative
behavior for the CFT dual of excitations near an AdS$_4$ brane.
However, we find that the CFT shell becomes infinitely large when the
bulk excitation is still only a finite distance from the brane.  This
occurs precisely when the excitation reaches the boundary of the
holographic domain calculated in Sec.~\ref{sec-domains}.

This provides a beautiful consistency check, and it answers the
question raised earlier.  When the particle exits the brane's
holographic domain, the holographic image disappears from the AdS$_4$
brane by moving off to spatial infinity.  In Sec.~\ref{sec-disc}, we
consider the global structure of the AdS$_4$/AdS$_5$ system, which
suggests an intriguing interpretation (Sec.~\ref{sec-interpret}).
Unlike the RS model, the AdS$_4$/AdS$_5$ system retains half of the
boundary of AdS$_5$.  Its holographic domain is evidently the
remaining portion of the bulk that was not covered by the brane's
holographic domain.  As a particle crosses back and forth between
domains, the holographic image must cross between the CFT on the
brane, and the CFT on the AdS$_5$ half-boundary.  We also discuss
implications for the break-down of four-dimensional gravity on the
brane.

\section{Branes and space-like holography in AdS$_5$}
\label{sec-branes}

In this section, we present metrics for flat and AdS$_4$ branes in
AdS$_5$.  In each case we relate brane volumes to bulk volumes.  We
examine the compatibility of this space-like relation with independent
results for the strength of four-dimensional gravity, and with the
holographic principle.  In the RS case, we find no contradictions.  In
the AdS$_4$ case, we find a discrepancy of the space-like relation
with the finiteness of the four-dimensional Planck mass, and we expose
a violation of the holographic principle.

\subsection{RS brane}
\label{sec-rs}

Newton's constant, $G_{\rm N}=M_4^{-2}$, need not vanish even in the
presence of non-compact extra dimensions. Consider five-dimensional
gravity%
\footnote{We use the conventions of Wald~\cite{Wald} for the
 metric signature and the Ricci scalar.  The quantity
 ``$\lambda$'' in~\cite{KarRan00} corresponds to $4\pi T/ M_5^3$
 in our notation.}
with a negative cosmological constant, coupled to a three-brane of
tension $T$:
\begin{equation}
S = \frac{M_5^3}{16\pi} \int d^5x \sqrt{-g} \left(
R_5 + \frac{12}{\ell^2} \right)- T \int d^4x\, dr\, 
\delta(r-c) \sqrt{-\det g_{ij}}.
\end{equation}
The Randall-Sundrum model~\cite{RS} embeds a 3+1 dimensional Minkowski
brane into portions of 4+1 dimensional Anti-de~Sitter space (AdS$_5$).
This requires that the tension takes the value
\begin{equation}
T=\frac{3M_5^3}{4\pi\ell}.
\label{eq-tension}
\end{equation}

The solution can be written as a warped product metric,
\begin{equation}
{ds^2\over \ell^2} = e^{2r}
\left( -dt^2 + d\rho^2 +\sin^2\rho\, d\Omega_2^2 \right) + dr^2,
\label{eq-minkwarp}
\end{equation}
where $\ell$ is the curvature scale of the AdS$_5$ space-time.  The
brane resides at an arbitrary fixed value of the warp coordinate,
$r=c$.  (Conventionally $r=0$ is chosen, but with a view to the
AdS$_4$ case we will find instructive to leave $c$ unspecified.)  The
portion of the above metric with $r>c$ is removed, and one can think
of the brane as a ridge at which the remaining portion of the AdS$_5$
Poincar\'e patch, $-\infty<r<c$, is matched to another copy of itself.
We will consider the two copies to be identified under
${\mathbb{Z}}_2$.

Let us pick a finite three-dimensional spatial region $V_3$ at a fixed
time $t$ on the brane (Fig.~\ref{fig-avav}).
\EPSFIGURE{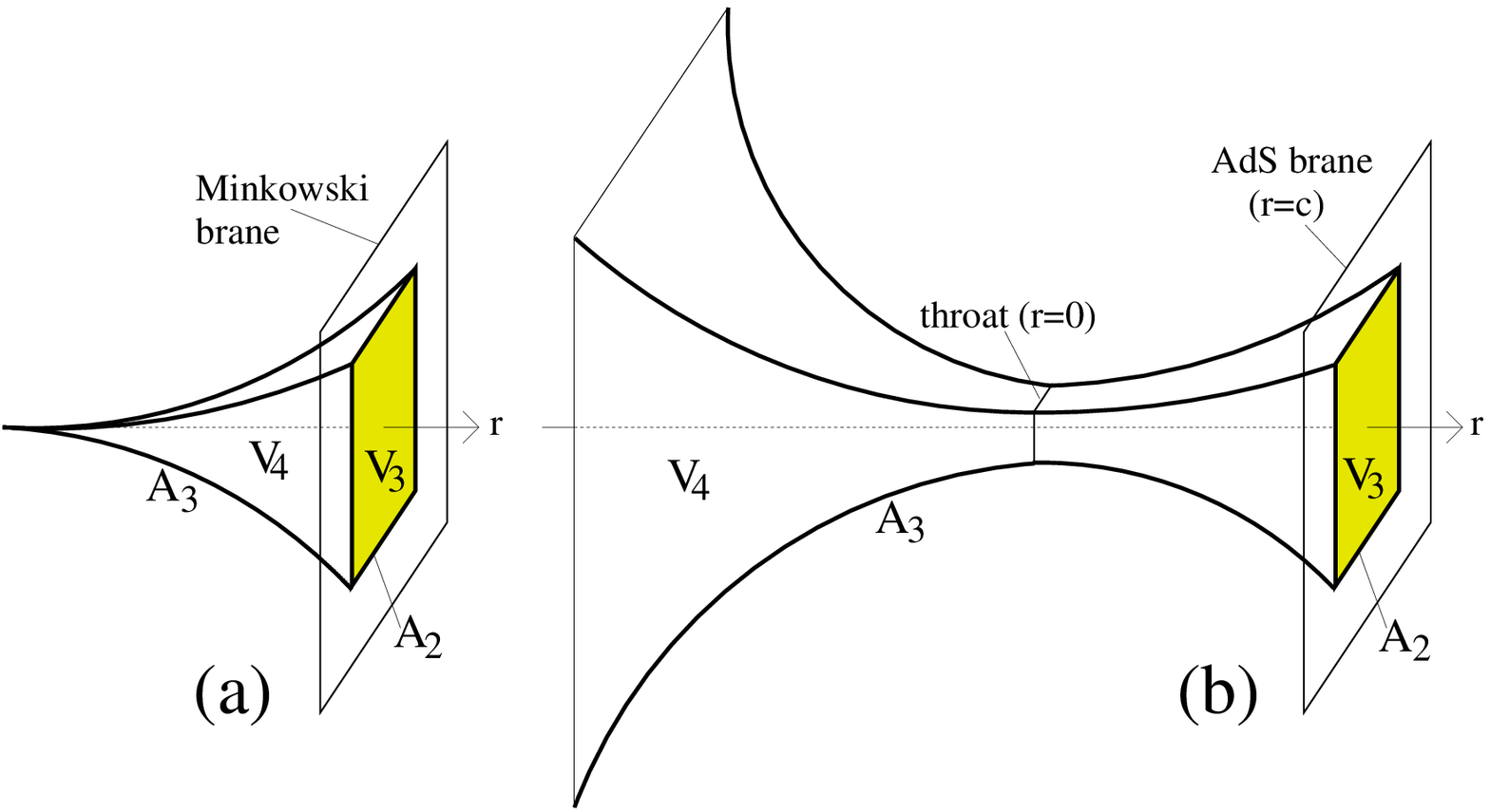,width=13cm}%
{Branes at $r=c$ in a warped compactification of AdS$_5$; the time
direction is suppressed. (a) A Minkowski brane has warp factor $e^r$.
A finite volume $V_3$ on the brane corresponds to a finite volume
$V_4$ in the bulk; similarly for areas $A_2$ and $A_3$.  (b) An
AdS$_4$ brane has warp factor $\cosh r$.  Beyond the throat ($r=0$)
the bulk area and volume diverge.  The physics in the infinite region
$V_4$ cannot be holographically represented by CFT excitations in
$V_3$.
\label{fig-avav}}
We associate to it a four-dimensional spatial volume $V_4$ by warped
multiplication, i.e., by lifting the restriction $r=c$.%
\footnote{Warped multiplication is clearly the most natural way to
associate brane and bulk volumes at equal time.  Nevertheless, one
should keep in mind that it is based on a coordinate choice.  Strictly
speaking, the space-like relation between brane and bulk volumes is
therefore not even uniquely defined.  By contrast, the covariant
relation via light-sheets is unambiguous
(Sec.~\ref{sec-ls}).}
Even though the $r$ coordinate extends an infinite proper distance
away from the brane, this bulk volume will be finite:
\begin{equation}
V_4 = \int_{-\infty}^c \ell dr\, V_3 e^{r-c} = V_3 \ell.
\label{eq-v4v3}
\end{equation}

In standard Kaluza-Klein compactification, the size of the extra
dimension determines the four-dimensional Planck scale in terms of the
five-dimensional one.  The relation can be written in the form
\begin{equation}
\frac{M_4^2}{M_5^3} = \frac{V_4}{V_3},
\label{eq-m4m5}
\end{equation}
suggestive of the dilution of the gravitational field in the extra
dimension.  One might expect that (\ref{eq-m4m5}) also applies to
warped compactifications such as the RS model.  Then (\ref{eq-v4v3})
implies the relation
\begin{equation}
M_4^2 = M_5^3 \ell
\end{equation}
between Planck masses in the RS model.  This is indeed the relation
found in~\cite{RS} by analysis of graviton modes.  We conclude that
the association of bulk and brane regions by warped multiplication is
consistent with the independently computed strength of
four-dimensional gravity on the brane.

Space-like holography demands that CFT states with support on $V_3$
encode all bulk information in $V_4$, requiring in particular that
\begin{equation}
S_5(V_4) \leq {V_3 M_5^3 \over 4}.  
\label{eq-s5v3} 
\end{equation}
The entropy $S_5$ in the bulk region $V_4$ must not exceed the
``area'' $V_3$, in five-dimensional Planck units.

Since $V_3$ does not enclose $V_4$, we cannot use the space-like
condition to argue that (\ref{eq-s5v3}) follows from the covariant
holographic principle.  Indeed, after complementing the geometry in
Fig.~\ref{fig-avav}a with its mirror image across the brane, we
recognize that $V_3$ is not even a portion of the boundary of
$V_4$. The actual boundary of $V_4$, $A_3$, resides mostly in the
bulk.  Since $A_3$ is a closed surface around $V_4$ (we may neglect
the point at infinity), it satisfies the space-like condition.  It
follows that
\begin{equation}
S_5(V_4) \leq \frac{A_3 M_5^3}{4}.
\label{eq-s5}
\end{equation}
This inequality is stronger than (\ref{eq-s5v3}), as we show next.

Let $A_2$ be the intersection of $A_3$ with the brane.  Then $A_3$ is
related to $A_2$ by warped multiplication:
\begin{equation}
A_3 = \int_{-\infty}^c \ell dr\, A_2 e^{r-c} = A_2 \ell.
\label{eq-a3a2}
\end{equation}
Note that $A_2$ is the boundary of the brane region $V_3$.  Moreover,
we may assume that $V_3$ has no dimensions smaller than the CFT
cut-off scale, $\ell$; otherwise, CFT states cannot be localized to
$V_3$.  It follows that $V_3 > A_2 \ell = A_3$.  With (\ref{eq-s5})
this implies the inequality (\ref{eq-s5v3}).  We conclude that
space-like holography, as defined by projection along the warp
direction $r$, is consistent with the holographic entropy bound in the
RS model.

We note parenthetically that (\ref{eq-a3a2}) implies an interesting
statement, which however will not be needed later.  Since the
four-dimensional Planck mass is finite, the holographic principle must
also apply to the four-dimensional theory on the brane:
\begin{equation}
S_4(V_3) \leq \frac{A_2 M_4^2}{4}.
\label{eq-s4}
\end{equation}
Moreover, we assume that the CFT encodes all information in the bulk,
i.e., $S_4(V_3) = S_5(V_4)$.  Then (\ref{eq-a3a2}) and (\ref{eq-m4m5})
imply the equivalence of (\ref{eq-s5}) and (\ref{eq-s4}).  The
holographic entropy bound is equivalent in the four- and the
five-dimensional theory, as it should be.

\subsection{AdS branes and the break-down of space-like holography}
\label{sec-ads}

The de-tuning of the brane tension (\ref{eq-tension}) leads to
generalizations of the RS
model~\cite{DewFre99,Kal99,KimKim99,Nih99,KarRan00} in which the brane
is no longer flat.  Instead, the brane resides at constant $r$ in a
warped slicing of an AdS$_5$ bulk into dS$_4$ or AdS$_4$ slices.  The
latter case is of particular interest because the warp factor, $\cosh
r$, fails to be monotonic:
\begin{equation}
{ds^2\over \ell^2} = \cosh^2 r \left( -\cosh^2 \rho\, 
dt^2 + d\rho^2 +\sinh^2\rho\, d\Omega_2^2 \right) + dr^2.
\label{eq-adswarp}
\end{equation}
An AdS$_4$ brane of positive tension $T = \frac{3M_5^3}{4\pi\ell}\tanh
c$ resides at $r=c>0$.  The portion of AdS$_5$ with $r>c$ is removed.
We refer to the AdS$_4$ hypersurface $r=0$ as the {\em throat} of the
warped geometry, because the warp factor has a minimum there.  Near
the brane, between $r=c$ and the throat, the spatial bulk geometry
shrinks away from the brane.  This mimics the $e^{r}$ behavior of the
warp factor in the RS model.

Beyond the throat, from $r=0$ to $r=-\infty$, the warp factor
increases without bound (Fig.~\ref{fig-avav}).  Hence, the bulk volume
associated with a given volume on the brane diverges:
\begin{equation}
V_4 = \int_{-\infty}^c \ell dr\, V_3 {\cosh r \over \cosh c}
 \rightarrow \infty.
\label{eq-v4v3ads}
\end{equation}

Reasoning analogous to (\ref{eq-m4m5}) would suggest that the
four-dimensional Planck mass should diverge.  A mode analysis has
shown, however, that gravity nevertheless localizes on the
brane~\cite{KarRan00}, with $M_4^2 \approx M_5^3 \ell$.  The covariant
methods developed in Sec.~\ref{sec-diamonds} resolve this
contradiction, as we discuss at the end of Sec.~\ref{sec-disc}.  We
will find that gravity looks four-dimensional only for a limited time.
The relevant time scale is the AdS$_4$ curvature radius
\begin{equation}
\ell_4 = \ell \cosh c,
\end{equation}
which can be made parametrically larger than the AdS$_5$ radius,
$\ell$, by choosing $c\geq 1$.%
\footnote{This corresponds to an appropriate choice of brane tension,
which we treat as an adjustable parameter of the theory.  We will not
be concerned with the microscopic origin of these scales because that
problem seems unrelated to the questions addressed here.
See~\cite{KarRan01a,KarRan01b} for a recent approach.}

The space-like holographic principle, inequality (\ref{eq-s5v3}), is
clearly violated in the AdS$_4$/AdS$_5$ system.  The ``area'' $V_3$ is
finite, but the divergent bulk volume $V_4$ can contain arbitrarily
large systems with entropy
\begin{equation}
S_5(V_4) \gg V_3 M_5^3.
\label{eq-prdox}
\end{equation}
Hence, the bulk physics cannot be equivalent to any field theory on
the AdS$_4$ brane.  This appears to conflict with the effective
compactification to four dimensions indicated by the mode analysis
of~\cite{KarRan00}.

Next we review the covariant form of the holographic principle.  In
Sec.~\ref{sec-domains}, we will show how it resolves the paradox
(\ref{eq-prdox}).

\section{Light-sheet holography}
\label{sec-ls}

The apparent contradictions we have found originate from our attempt
to relate brane and bulk volumes in a naive, space-like manner.  In
this section we review the covariant entropy bound, a more precise and
general formulation of the holographic principle.  We describe how to
calculate the extent of the holographic region bordering on a given
area.  We also discuss how to apply the covariant formalism to a brane
of codimension 1.

\subsection{Covariant entropy bound}
\label{sec-ls1}

The covariant formulation of the holographic
principle~\cite{Bou99b,Bou99c,Bou99d} associates bulk regions to
areas.  In $D$ space-time dimensions, by an area we mean a surface of
$D-2$ spatial dimensions, i.e., {\em not}\/ the history of a $D-2$
surface.  

There are four families of light-rays orthogonal to any area $A$, as
shown in Fig.~\ref{fig-cebwarp}: 
\EPSFIGURE{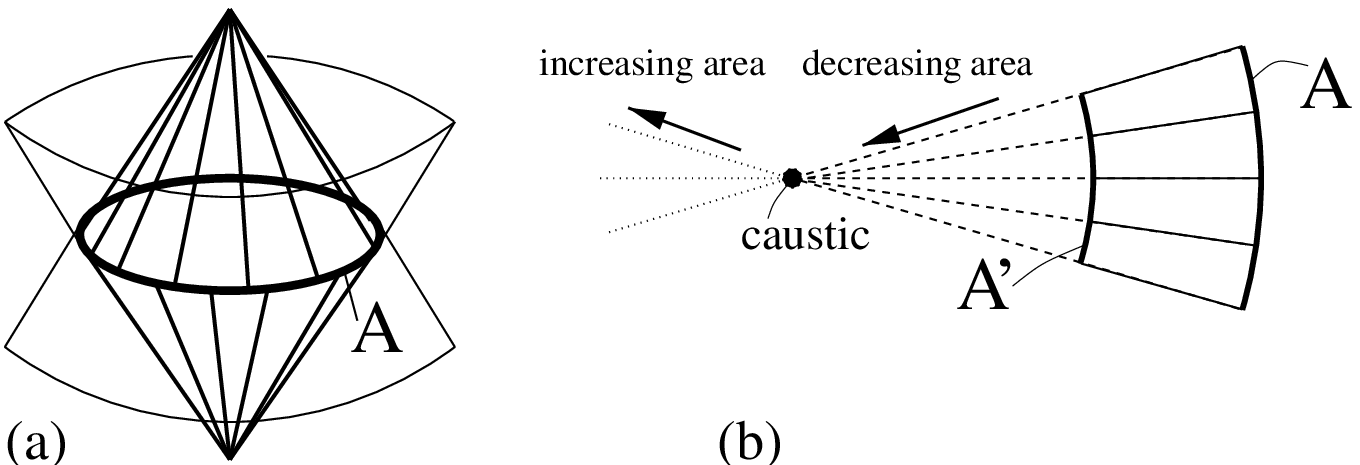,width=12cm}%
{(a) A space-time picture of the two light-sheets of a spherical area
$A$.  The entropy on either cone is bounded by $A$.  The other two
light-like directions are the skirts drawn in thin outline.  Their
cross-sectional area is increasing, so the entropy of the skirts is
not related to $A$.  (b) [Here the time direction is suppressed.]  The
requirement of decreasing area, $A'<A$, is a local condition.  When
neighboring light-rays intersect, the light-sheet must be terminated.
\label{fig-cebwarp}}
two to the past, and two to the future.  Consider one such family,
with affine parameter $\lambda$ along the light-rays.  The
cross-sectional area $A'(\lambda)$ spanned by the light-rays will
change as one follows the rays away from $A$.  By continuity,
$A'(\lambda)$ must be non-increasing for at least two of the four
families.  The hypersurfaces generated by the light-rays pointing in
any one of these ``shrinking'' directions are called {\em
light-sheets}.

Thus, light-sheets are null hypersurfaces of dimension $D-1$. They
would represent an equal-time hypersurface if we used light-cone
coordinates.  In fact, they represent a way of taking a snapshot of
matter systems.  The holographic principle states that the number of
degrees of freedom, and hence the entropy, on any light-sheet of $A$
is bounded by $A/4$ in Planck units:
\begin{equation}
S (\mbox{lightsheet of }A) \leq \frac{A M_D^{D-2}}{4},
\end{equation}
where $M_D$ is the Planck mass in the $D$-dimensional theory.

Formally, light-sheets are characterized by
\begin{equation}
\theta \leq 0.
\label{eq-thetacond}
\end{equation}
The expansion, $\theta$, of a family of light-rays, defined in
(\ref{eq-thetadef}) below, is a local measure of the rate of change of
the infinitesimal area element spanned by neighboring light-rays.  The
condition (\ref{eq-thetacond}) generalizes the usual notion of
``inside''.  It is well-defined where conventional notions fail, as in
closed or highly dynamical geometries, and for open surfaces.  For
convex closed surfaces in weakly gravitating regions of asymptotically
flat space, it reproduces the intuitive answer: Inside is where
infinity is not.  The two families that sweep through the enclosed
(``inside'') region, one past-directed and one future-directed, are
indeed contracting.  They form a past and a future light-cone ending
on $A$.  The light-rays going to the outside have growing
cross-sectional area and must be excluded from consideration
(Fig.~\ref{fig-cebwarp}).

The local condition (\ref{eq-thetacond}) also determines where
light-sheets must be terminated.  The light-rays may be followed only
as long as the expansion remains non-positive~\cite{Bou99b}.  If the
null energy condition,
\begin{equation}
T_{ab} k^a k^b \geq 0 \mbox{~~~for all null vectors~} k^a,
\end{equation}
is satisfied, the expansion can only become positive after diverging
to $-\infty$.  This happens at caustics, when neighboring light-rays
intersect.  Hence, light-sheets must be terminated at caustics, such
as the tips of the light-cones in Fig.~\ref{fig-cebwarp}.

\subsection{Finding caustics}
\label{sec-ls2}

In this subsection we show how to calculate the location of caustics
on a light-sheet of a given area $A$.

Let $A$ be a surface of $D-2$ spatial dimensions, parametrized by
coordinates $x^\alpha$, $\alpha=1,\ldots,D-2$.  Pick one of the four
families of light-rays that emanate from $A$ into the past and future
directions to either side of $A$.  Each light-ray satisfies the
equation for geodesics:
\begin{equation}
\frac{dk^a}{d\lambda} + \Gamma^a_{~bc} k^b k^c =0.
\label{eq-geodesic}
\end{equation}
where $\lambda$ is an affine parameter.  The tangent vector $k^a$ is
defined by
\begin{equation}
k^a = \frac{dx^a}{d\lambda}
\end{equation}
and satisfies the null condition $k^a k_a=0$.  

Let $l^a$ be the null vector field on $A$ that is orthogonal to $A$
and satisfies $k^a l_a=-2$.  (This means that $l^a$ has the same time
direction as $k^a$ and is tangent to the orthogonal light-rays
constructed on the other side of $A$.)  The induced $D-2$ dimensional
metric on the surface $A$ is given by
\begin{equation}
h_{ab} = g_{ab} + \frac{1}{2} \left(k_a l_b + k_b l_a\right).
\label{eq-induce}
\end{equation}
In a similar manner, an induced metric can be found for all other
cross-sections of the light-sheet.  The {\em null extrinsic curvature}
\begin{equation}
B_{ab} = h^c_{~a} h^d_{~b} \nabla_c k_d
\label{eq-bab}
\end{equation}
contains information about the expansion, $\theta$, shear,
$\sigma_{ab}$, and twist, $\omega_{ab}$, of the light-sheet:
\begin{eqnarray}
\label{eq-thetadef}
\theta &=& h^{ab} B_{ab}, \\
\sigma_{ab} &=& \frac{1}{2} \left( B_{ab}+B_{ba} \right) -
\frac{1}{D-2}\theta h_{ab}, \\
\omega_{ab} &=& \frac{1}{2} \left( B_{ab}-B_{ba} \right).
\end{eqnarray}

The Raychaudhuri equation describes the change of the expansion along
the light-rays:
\begin{equation}
\frac{d\theta}{d\lambda} = -\frac{1}{D-2} \theta^2 -
\sigma_{ab}\sigma^{ab} + \omega_{ab}\omega^{ab} - 8\pi G_{\rm N}
T_{ab} k^a k^b.
\label{eq-ray}
\end{equation}
For a surface-orthogonal family of light-rays, the twist vanishes.
The final term, $-T_{ab} k^a k^b$, will be non-positive if the null
energy condition is satisfied by matter.  Then the right hand side of
the Raychaudhuri equation is manifestly non-positive, and one can show
the following lemma: If the expansion at $\lambda=\lambda_0$ takes on
the non-zero value $\theta_0$, then there is a caustic (i.e.,
$|\theta| \rightarrow \infty$) somewhere between $\lambda_0$ and
$\lambda_0-\frac{D-2}{\theta_0}$.  This statement holds separately for
each light-ray in the family; in general, $\theta_0 =
\theta_0(x^\alpha)$.

Note that we can take $(\lambda, x^\alpha)$ to parameterize
simultaneously two of the four families of light-rays orthogonal to
$A$; one family has $\lambda>\lambda_0$ while the other has
$\lambda<\lambda_0$.  We have defined $\theta$ to describe the
expansion of the former family.  Since $\theta_0 \rightarrow
-\theta_0$ under $\lambda \rightarrow \lambda_0-\lambda$, we now see
explicitly that at least one of the pair forms a light-sheet.  For
$\theta_0 \neq 0$, the lemma shows that there will necessarily be a
caustic on the light-sheet.

To obtain the exact caustic position for a given matter distribution
$T_{ab}$, one would have to solve the Raychaudhuri equation and a
coupled evolution equation for the shear:
\begin{equation}
k^c \nabla_c \sigma_{ab} =
-\theta \sigma_{ab} + h^e_{~a} h^f_{~d} C_{cbef} k^c k^d.
\label{eq-shearev}
\end{equation}
The calculation simplifies under certain conditions, which will be
seen to hold in the example studied in Sec.~\ref{sec-domains}.
Suppose that the Weyl tensor, $C_{cbef}$, vanishes everywhere on the
light-sheet.  Suppose further that the shear vanishes initially:
$\sigma_{ab}(\lambda_0)=0$.  Then it follows from (\ref{eq-shearev})
that the shear vanishes everywhere on the lightsheet.  If the further
condition
\begin{equation}
\theta^2(\lambda_0) \gg G_{\rm N} T_{ab} k^a k^b
\label{eq-wimpy}
\end{equation}
holds for all $\lambda$, the effect of matter on the focusing is
negligible compared to the non-linear term, and (\ref{eq-ray})
becomes
\begin{equation}
\frac{d\theta}{d\lambda} = -\frac{1}{D-2} \theta^2.
\end{equation}
It follows that the caustics' location depends only on the initial
expansion and is given by
\begin{equation}
\lambda_{b}(x^\alpha) =\lambda_0 - \frac{D-2}{\theta_0(x^\alpha)}.
\label{eq-caustic}
\end{equation}
Generically the caustics form a surface of the same dimension as $A$.

Once the affine parameter at each caustic is known, its space-time
coordinates can be found by solving the null geodesic equation
(\ref{eq-geodesic}).

\subsection{Holographic domains}
\label{sec-ls3}

In the RS model and the generalizations we consider, the brane
occupies a time-like hypersurface $H_{D-1}$ of AdS$_5$.  But to
construct a light-sheet, we need to start from a spatial surface of
codimension 2.  Such surfaces can be obtained by slicing the brane
into surfaces of equal time, $V_{D-2}(t)$, as shown in
Fig.~\ref{fig-slicing}.  For each $V_{D-2}(t)$ we can construct a
light-sheet $L_{D-1}(t)$.  The resulting sequence of $(D-1)$
dimensional light-sheets will foliate a $D$-dimensional portion of the
bulk, the {\em holographic domain}\/ ${\cal D}(H_{D-1})$ associated
with the brane $H_{D-1}$.
\EPSFIGURE{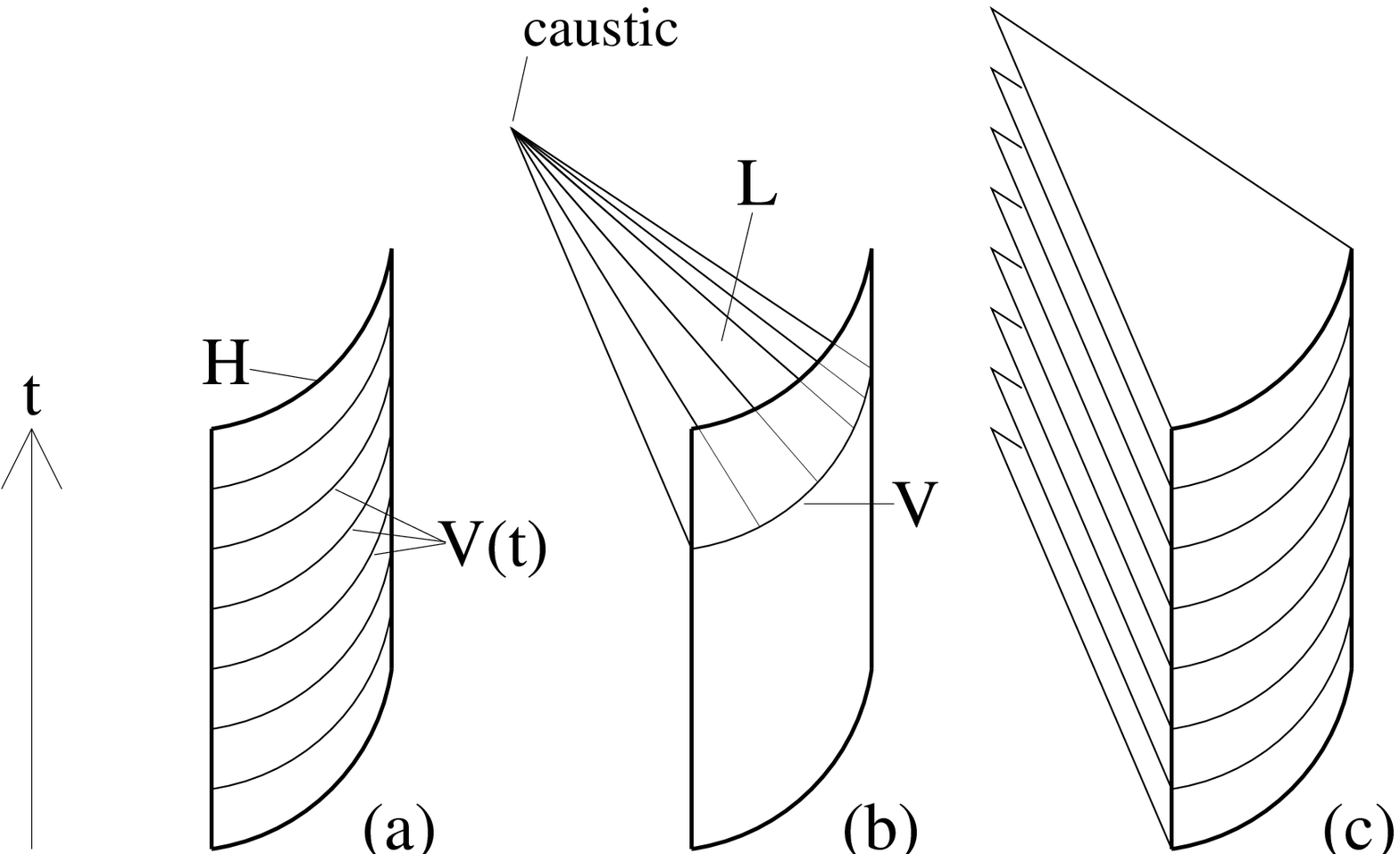,width=11cm}%
{Constructing the holographic domain ${\cal D}$ of a brane of
codimension 1, $H_{D-1}$.  (a) Slice $H_{D-1}$ into equal time
surfaces, $V_{D-2}(t)$ (b) Each $V_{D-2}(t)$ is an area from the bulk
point of view.  Construct its light-sheet, $L_{D-1}(t)$.  (c) The
light-sheets fill out a portion of the bulk, ${\cal D}(H_{D-1})$.
\label{fig-slicing}}

In general, the location of caustics will depend on the choice of
slicing, which we have not yet specified.  A different foliation
$V'_{D-2}(t')$ will have different light-sheets $L'_{D-2}(t')$, and
would yield a different answer for ${\cal D}(H_{D-1})$.  We also have
not specified which of the two allowed light-sheets should be
considered.  

Here we propose to resolve both ambiguities by demanding time reversal
invariance.  In the cases of interest the brane slices are {\em
normal}\/~\cite{Bou99b}, that is, they possess a past and a future
light-sheet going to the same side.  For a given slicing $V_{D-2}(t)$
let ${\cal D}^+[V_{D-2}(t)]$ (${\cal D}^-[V_{D-2}(t)]$) be the region
foliated by the future (past) light-sheets $L^+_{D-1}(t)$
($L^-_{D-1}(t)$).  One might declare, say, that the holographic domain
is always defined by past light-sheets (${\cal D}={\cal D}^-$).  But
then the brane would encode a different bulk region after time
reversal, even though the CFT is time-reversal invariant.

This motivates the demand that the slicing $V_{D-2}(t)$ be chosen so
that
\begin{equation}
{\cal D}^+[V_{D-2}(t)]=
{\cal D}^-[V_{D-2}(t)] \equiv {\cal D}(H_{D-1}).
\label{eq-pmcond}
\end{equation}
In other words, we require that it should not matter whether past or
future light-sheets are used to construct the holographic domain.%
\footnote{${\cal D}^+$ and ${\cal D}^-$ may contain the same
information even if they differ.  For example, suppose that ${\cal
D}^+={\cal D}^- - Q$, where the points in $Q$ are causally independent
of any points outside ${\cal D}^+$.  This motivates the refined
condition $D^\pm\{{\cal D}^+[V_{D-2}(t)]\}=D^\pm\{{\cal
D}^-[V_{D-2}(t)]\}$, where $D^\pm(U)$ denotes the union of the past
and future domains of dependence of the set $U$~\cite{HawEll}.  This
refinement will not play a role in the examples we study.}
We will not analyze here whether this condition is appropriate for
arbitrary brane and bulk metrics.  It is a reasonable condition for
the present purpose.

For AdS$_4$ and RS branes in AdS$_5$, we can find suitable slices
$V_{D-2}$ by demanding that the null extrinsic curvature of the
future-directed light-rays, $B_{ab}^+$, should be the same as that of
the past-directed light-rays, $B_{ab}^-$, at every point $x^\alpha$ on
$V_{D-2}$:
\begin{equation}
B_{ab}^+(x^\alpha) = B_{ab}^-(x^\alpha).
\end{equation}
This is equivalent to the condition that the slices $V_{D-2}(t)$ have
vanishing (ordinary) extrinsic curvature in $H$:
\begin{equation}
h_a^{~c}\, \nabla_c t_b =0.
\label{eq-extc}
\end{equation}
Here $t^c$ is the future-directed unit normal vector of $V_{D-2}(t)$
in $H_{D-1}$.  

For $H=\,$AdS$_4$, (\ref{eq-extc}) is uniquely satisfied by the
hyperbolic space
\begin{equation}
{ds_3^2\over \ell_4^2} = d\rho^2 +\sinh^2\rho\, d\Omega_2^2.
\label{eq-hyper}
\end{equation}
Any complete foliation of $H$ by such spaces will satisfy the
condition (\ref{eq-pmcond}) and will give the same answer for $\cal
D$.  The simplest example is given by the ``global coordinates'' of
AdS$_4$,
\begin{equation}
{ds_4^2\over \ell_4^2} = -\cosh^2 \rho\, dt^2 + d\rho^2 +\sinh^2\rho\,
d\Omega_2^2.
\label{eq-global}
\end{equation}
We have chosen the AdS$_5$ metric (\ref{eq-adswarp}) to conform to
these coordinates. 

For $H=\,$M$^4$, (\ref{eq-extc}) is satisfied by the flat ${\mathbb
R}^3$ slices arising in the metric (\ref{eq-minkwarp}).

\section{Holographic domains in AdS$_5$}
\label{sec-domains}

In this section we locate the caustics of light-sheets of a given
brane in AdS$_5$.  This determines the extent of the brane's
holographic domain in the bulk.  We describe the calculation in detail
for an AdS$_4$ brane and also state the result for the RS brane.

The metric of an AdS$_4$ brane immersed in AdS$_5$ was given in
(\ref{eq-adswarp}):
\begin{equation}
{ds^2\over \ell^2} = \cosh^2 r \left( -\cosh^2 \rho\, 
dt^2 + d\rho^2 +\sinh^2\rho\, d\Omega_2^2 \right) + dr^2.
\label{eq-adswarp2}
\end{equation}
The time slicing induced on the brane, at $r=c$, obeys the condition
(\ref{eq-extc}).  Because the metric is static, it will suffice to
consider the future-directed light-sheet of one time-slice, $t=0$.  We
will compute the caustic location, $r=b$.  The holographic domain
associated to the brane will be characterized by $b\leq r\leq c$.

Following the general method described in Sec.~\ref{sec-ls}, we will
first determine the caustic position in terms of the affine parameter
$\lambda$; then we will find $r(\lambda)$.

The null vector fields $k^a$ and $l^a$, defined by
\begin{equation}
k^t = \frac{1}{\cosh r \cosh \rho},~~k^r ={-1},~~
k^\rho=k^\theta=k^\phi=0
\label{eq-ka}
\end{equation}
and
\begin{equation}
l^t = \frac{1}{\cosh r \cosh \rho},~~l^r ={1},~~
l^\rho=l^\theta=l^\phi=0
\label{eq-la}
\end{equation}
satisfy $k^a l_a=-2$ and are orthogonal to the brane slice defined by
$r=c$, $t=0$.  At this slice, the vector field $k^a$ ($l^a$) coincides
with the tangent vector field of the future-directed orthogonal
light-rays going in the negative (positive) $r$ direction.  Keeping in
mind that the region $r>c$ is removed from (\ref{eq-adswarp2}) and
replaced by a second copy of the geometry, we need only consider the
light-rays with tangent vector $k^a$, which we will find to have
negative expansion.  Note that (\ref{eq-ka}) and (\ref{eq-la}) are not
necessarily orthogonal to any other cross-sections of the light-sheet,
but they will allow the computation of the initial expansion and
shear.

According to (\ref{eq-induce}), the induced metric on the brane
slice is given by
\begin{equation}
h^a_{~b} = {\rm diag}(0,0,1,1,1).
\end{equation}
With (\ref{eq-bab}) one finds
\begin{equation}
B^a_{~b} = -\tanh c\ {\rm diag}(0,0,1,1,1).
\end{equation}
Hence, the initial expansion of the light-sheet is independent of the
position along the brane slice:
\begin{equation}
\theta_0 = -3 \tanh c
\end{equation}
As required, the twist vanishes.  We note that the initial shear also
vanishes.  Moreover, the Weyl tensor $C_{cbad}$ vanishes in vacuum
AdS$_5$.  We conclude that the shear vanishes everywhere on the
light-sheet.  For $c > 1$, the initial expansion is of order unity, so
that condition (\ref{eq-wimpy}) is guaranteed to hold in the weak
field limit.  Hence, the location of the caustic is determined by the
initial expansion alone.  The value of the affine parameter at the
brane may be set to zero.  Then (\ref{eq-caustic}) yields
\begin{equation}
\lambda_b = \coth c.
\end{equation}

The geodesic equation (\ref{eq-geodesic}) for the $r$ coordinate is
\begin{equation}
\frac{d^2r}{d\lambda^2} + \sinh r \cosh r \left[ \cosh^2 \rho \left(
\frac{dt}{d\lambda} \right)^2 - \left( \frac{d\rho}{d\lambda}
\right)^2 \right] = 0.
\end{equation}
The tangent vector is null:
\begin{equation}
\cosh^2 \rho \left( \frac{dt}{d\lambda} \right)^2 
-\frac{1}{\cosh^2 r} \left( \frac{dr}{d\lambda} \right)^2
- \left( \frac{d\rho}{d\lambda} \right)^2=0.
\end{equation}
Hence, the $r$ equation simplifies to
\begin{equation}
\frac{d^2r}{d\lambda^2} + \tanh r \left( \frac{dr}{d\lambda} \right)^2
=0,
\end{equation}
with general solution
\begin{equation}
\lambda(r) = C_1 \sinh r + C_2.
\end{equation}
At $r=c$, one has $\lambda= 0$ and $k^r = \frac{dr}{d\lambda}=-1$,
which fixes the constants:
\begin{equation}
\lambda(r) = \frac{\sinh c - \sinh r}{\cosh c}.
\end{equation}

The caustic lies at $r=b$ with $\lambda(b) =\lambda_b$.  Hence,
\begin{equation}
\sinh b = \frac{-1}{\sinh c}.
\label{eq-cau}
\end{equation}
The calculation involved the assumption $c \geq 1$, which is also
required for the localization of gravity over a significant
scale~\cite{KarRan00}.  

\EPSFIGURE{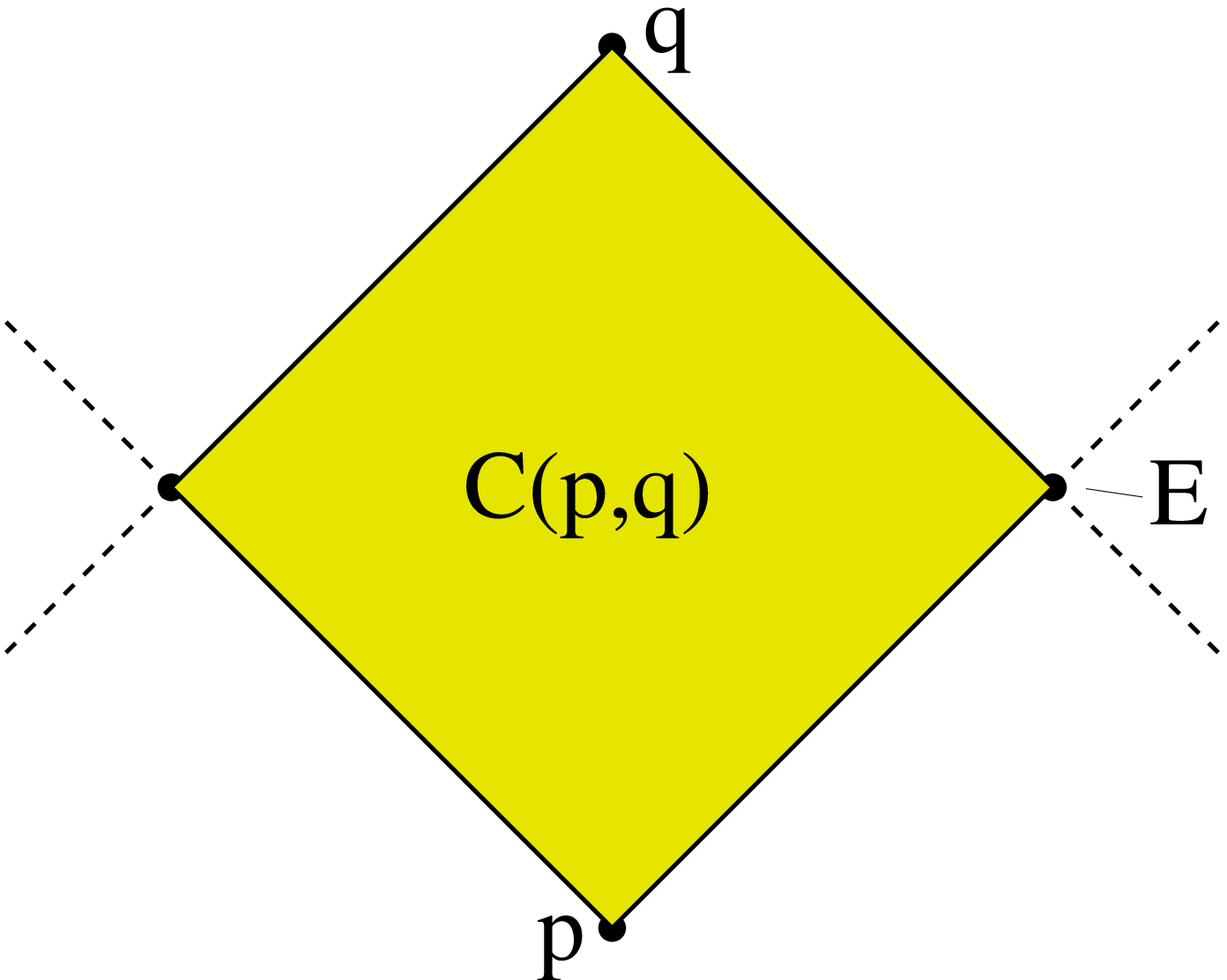,width=6cm}%
{A causal diamond $C(p,q)$ is the space-time region probed by an
experiment beginning at $p$ and ending at $q$.
\label{fig-cdwarp}}%
Notice that we have shown that each light-ray in the light-sheet
terminates at $r=b\approx -2 e^{-c}$, just beyond the throat of the
warp factor.  We conclude that the holographic domain ${\cal D}(H)$ of
an AdS$_4$ brane $H$ at $r=c>1$ is the bulk region
\begin{equation}
b<r<c
\label{eq-range}
\end{equation} 
in the coordinates (\ref{eq-adswarp}).%
\footnote{
We have shown only that ${\cal D}(H)$ is contained in the region
(\ref{eq-range}).  By considering a past light-cone from $r=b$,
$\rho=0$, whose intersection with the hypersurface $r=b$ meets $H$ at
their common spatial infinity, one can show that ${\cal D}(H)$ is in
fact equal to the region $b<r<c$.}
The infinite region $r<b$ is not included on the light-sheet.  This
eliminates the possibility of violating the holographic principle with
large entropic systems in the region behind the throat.

An analogous but simpler calculation of light-sheets of the RS brane
shows that their caustics lie on the Poincar\'e horizon.  Hence, the
holographic domain of the RS brane is the surviving portion of the
Poincar\'e patch, i.e., the portion of AdS$_5$ satisfying
\begin{equation}
r<c
\label{eq-minkrange}
\end{equation} 
in the coordinates (\ref{eq-minkwarp}).

\section{Holographic images from causal diamonds}
\label{sec-diamonds}

Bulk physics taking place within a brane's holographic domain ${\cal
D}$ should have a holographic dual in terms of CFT excitations on the
brane; bulk physics outside ${\cal D}$ should not.  This raises a
sharp question, which we study in the next two sections.  How does the
holographic image disappear from the brane when a bulk particle exits
from ${\cal D}$?

In this section, we give a simple construction that allows us to
locate on the brane the CFT exitations dual to a bulk particle.  We
will apply this construction to the RS and KR models in the following
section.  Our construction is general and, in certain limits, can also
be applied in the AdS/CFT correspondence (unmodified by branes).%
\footnote{Causal diamonds have previously arisen in connection with
the AdS/CFT correspondence in Refs.~\cite{Reh99a,Reh99b,Reh00}.  We
would like to thank S.~Ross for bringing those references to our
attention.  An important early investigation of causality in the
AdS/CFT correspondence~\cite{HorItz99} was restricted to a set of
questions that did not require the use of causal diamonds.}

Consider a pointlike observer in an arbitrary space-time.  Suppose the
observer performs an experiment that begins at a point $p$ on his
world-line and ends at some later point $q$ (Fig.~\ref{fig-cdwarp}).
By causality, this experiment cannot probe points that lie outside the
past light-cone of $q$, because light does not have enough time to
reach the observer.  Moreover, no points outside the future light-cone
of $p$ can be probed.  Hence, the experiment can be associated to the
{\em causal diamond\/}~\cite{Bou00b}, $C(p,q)$, defined as the
space-time region that lies both in the future of $p$ and in the past
of $q$.  It is bounded by light-cones of $p$ and $q$, whose
intersection we call the {\em edge}, $E(p,q)$.

Any experiment beginning at $p$ and ending at $q$ can only learn
directly about physics taking place in the space-time region
$C(p,q)$. Indirectly, of course, inferences can be made about the
previous history of particles that entered the causal diamond from the
past.  But for the purpose of estimating the amount of information
that can be gained in such an experiment, it suffices to know the
maximum entropy that can be contained within $C(p,q)$.
\EPSFIGURE{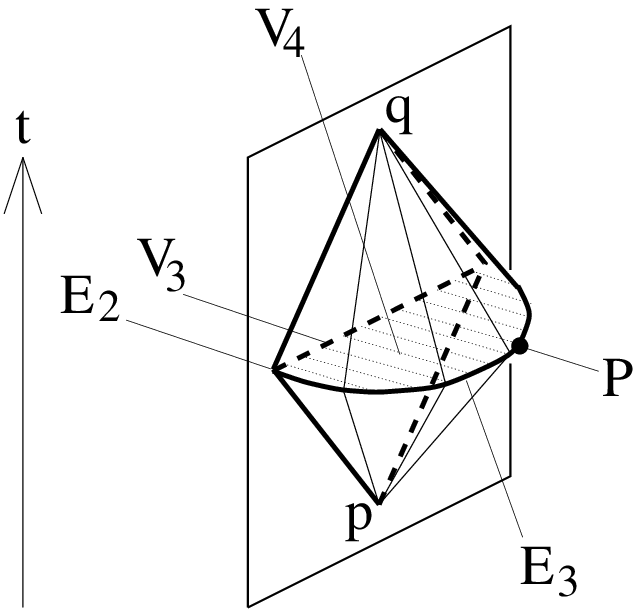,width=6cm}%
{Causal diamond barely containing a bulk event at $P$.  By causality,
the holographic image of $P$ has support on the intersection of the
edge with the brane.  Hence, it is shell of CFT energy whose radius
grows with the distance of $P$ from the brane.
\label{fig-bbdiamond}}

An observer trapped on a brane in AdS$_5$ has two ways of describing
certain experiments.  A probe of the state of the holographic CFT on
the brane can alternatively be described as an excursion into the
bulk.  These languages must be equivalent if the holographic duality
is faithful.  By causality, an experiment of finite duration cannot
probe bulk regions arbitrarily far from the brane.  It follows that
the CFT will not contain information about distant bulk regions when
probed at a sufficiently short distance scale.

More specifically, given a five-dimensional causal diamond $C_5(p,q)$
with $p$ and $q$ on the brane, one can associate to it a
four-dimensional causal diamond $C_4(p,q)$, obtained by intersecting
$C_5(p,q)$ with the brane (Fig.~\ref{fig-bbdiamond}).  Let $V_4(p,q)$
be a bulk volume bounded by the edge of $C_5(p,q)$.  Let $V_3(p,q)$ be
the intersection of $V_4(p,q)$ with the brane.  Hence, $V_3(p,q)$ is a
brane volume bounded by $E_2(p,q)$, the edge of $C_4(p,q)$.  By
causality, excitations of the CFT with support on $V_3(p,q)$ should
encode no more nor less than the bulk physics taking place in the
region $V_4(p,q)$.

We are interested specifically in the support ${\cal S}(P)$ of the CFT
state dual to a pointlike excitation in the bulk.  This can be
constructed as follows.  Given an excitation at a point $P$ off the
brane, find points $p$ and $q$ on the brane such that $C_5(p,q)$
contains $P$ but does not contain points farther from the brane than
$P$.  By causality, the CFT dual to the event at $P$ must have support
only on $V_3(p,q)$:
\begin{equation}
{\cal S}(P) \subset V_3(p,q).
\label{eq-high}
\end{equation}
Now consider $C_5(p',q')$, where $p'$ ($q'$) are brane events
infinitesimally later (earlier) than $p$ ($q$).  This causal diamond
is nested just within $C_5(p,q)$ and barely fails to include $P$.  In
other words, the $(p',q')$ experiment fails to detect $P$ in the bulk.
Then causality demands that it must also fail to probe the holographic
image of $P$ on the brane.  The $(p',q')$ experiment probes the brane
volume $V_3(p',q')$.  It follows that the CFT dual to $P$ has no
support on $V_3(p',q')$:
\begin{equation}
{\cal S}(P) \cap V_3(p',q') = \emptyset.
\label{eq-low}
\end{equation}
Combining (\ref{eq-high}) and (\ref{eq-low}), we conclude that 
\begin{equation}
{\cal S}(P) \subset V_3(p,q)-V_3(p',q') = E_2(p,q).
\end{equation}

Physically, the edge $E_2(p,q)$ is the distance scale explored on the
brane in an experiment that probes as far as to the point $P$ into the
bulk.  The CFT excitation dual to an event $P$ in the bulk (the {\em
holographic image}\/ of $P$) has support on the edge of the causal
diamond $C_4(p,q)$.

\section{Evanescence of CFT shells}
\label{sec-evanescence}

In this section we check the consistency of our calculation of the
holographic domain, Eqs.~(\ref{eq-range}) and (\ref{eq-minkrange}).
Suppose a local bulk excitation moves away from the brane and
eventually exits the holographic domain of the brane.  We would like
to verify that the holographic image disappears from the brane at this
moment.  For this purpose we apply the method of causal diamonds to
locate holographic images for RS and AdS$_4$ branes in AdS$_5$.  We
will show that the image moves off to infinity as the corresponding
bulk source exits the brane's holographic domain.

Recall from the metrics (\ref{eq-minkwarp}) and (\ref{eq-adswarp})
that $\rho$ is a radial coordinate on the brane in both cases.  The
coordinate $r$ parametrizes the warped direction away from the brane.
The brane resides at $r=c$.  We consider a bulk event $P$ at
$r=r_P<c$.  The translation symmetries in the brane directions allow
us to take $P$ to be at $\rho=0$, $t=0$.  

We must find a causal diamond $C_5(p,q)$ that contains $P$ and
contains no points farther from the brane than $P$.  This does not
completely fix $p$ and $q$.  It implies only that they will be at the
same value of $\rho$ and at antipodal angles on the sphere.  The boost
symmetry of the brane, however, allows us to take $p$ and $q$ to
reside at $\rho=0$.  We will determine their time coordinates, $\mp
\Delta t(r_P)$, by computing the time coordinate when a past (future)
light-ray from $P$ along $\rho=0$ reaches the brane at $r=c$.

The holographic image of $P$ lies on the edge of the causal diamond
$C_4(p,q)$.  The edge, $E_2(p,q)$, consists of the outermost points on
the brane reached by light-rays that start at $p$ and are reflected
back to $q$.  Hence, $E_2(p,q)$ is a shell of radius $\rho(r_P)$,
where $\rho(r_P)$ is the coordinate distance traveled by a light-ray
on the brane in the coordinate time $\Delta t(r_P)$.

\subsection{RS brane}
\label{sec-rsc}

In the RS model, described by the metric~(\ref{eq-minkwarp}), a bulk
light-ray between $P$ and $q$ obeys
\begin{equation}
\int_0^{\Delta t(r_P)} dt = \int_{r_P}^c e^{-r} dr.
\end{equation}
An analogous equation governs the light-ray from $p$ to $P$.
Solving for $\Delta t(r_P)$ one finds
\begin{equation}
\Delta t(r_P) = e^{-r_P} - e^{-c}.
\label{eq-trc1}
\end{equation}

Next we consider the light-rays on the brane that radiate out from $p$
and reflect on the edge $E_2(p,q)$ so as to focus at $q$.  The rays
returning from $E_2(p,q)$ to $q$ obey
\begin{equation}
\int_0^{\rho(r_P)} dr = \int_0^{\Delta t(r_P)} dt,
\end{equation}
with a similar equation holding for the outgoing rays.  It follows that
\begin{equation}
\rho(r_P) = \Delta t(r_P).
\label{eq-trc2}
\end{equation}

By inserting (\ref{eq-trc1}) into (\ref{eq-trc2}) we obtain the
desired relation between the position of a bulk event, $r_P$ and the
radius of its holographic image, $\rho(r_P)$:
\begin{equation}
\rho(r_P)=e^{-r_P}-e^{-c}~~~\mbox{(RS brane)}.
\label{eq-rsrelsize}
\end{equation}

This result, obtained only from considerations of causality, is
consistent with more elaborate studies of freely falling particles in
AdS$_5$~\cite{GidKat00b,GreRub00}.  A particle at $r_P$ in the bulk
corresponds holographically to CFT energy momentum localized on a
shell of radius $\rho(r_P)$.  As the particle moves deeper into the
bulk, the shell expands.  (We are working in the geometrical optics
approximation.  One would expect the width of the shell to be smeared
by a characteristic wavelength no shorter than the ultraviolet cutoff
of the CFT.  Indeed, for a particle at rest when released from the
brane, the width was found to be of order the AdS$_5$ radius, $\ell$,
in~\cite{GidKat00b}.)  By repeating the above procedure for every
point on a time-like bulk geodesic one finds that the shell expands
with constant acceleration.  Hence its world tube begins and ends on
the null infinities of Minkowski space, ${\cal I}^-$ and ${\cal I}^+$.

This sheds light on an interesting aspect of the AdS/CFT
correspondence.  Recall that the holographic domain of a Minkowski
brane, (\ref{eq-minkrange}), is (the surviving portion of) the
Poincar\'e region of AdS$_5$.  In the full AdS/CFT correspondence, a
similar relation holds between the (whole) Poincar\'e region and its
conformal boundary.  One might be concerned about the incompleteness
of the Poincar\'e region.  In the bulk, particles can enter through
the past Poincar\'e horizon and disappear behind the future Poincar\'e
horizon.  Of course this is just a consequence of using a non-global
coordinate system and does not imply loss of unitarity.  But the
Poincar\'e region is holographically dual to a field theory on a {\em
complete} space-time, four-dimensional Minkowski space.  How can the
dual theory handle the loss and gain of excitations and yet be
unitary?  

As a particle approaches the Poincar\'e horizon ($r_P\rightarrow
-\infty$) the shell on the brane becomes infinitely large at a rate
approaching the speed of light.  By energy conservation, it will also
become infinitely dilute.  This process takes an infinite proper time
for a brane observer.  At the classical level, we thus understand how
the holographic image disappears when a particle exits the holographic
domain of the RS brane.  The boundary of the Poincar\'e region is
mapped to the conformal boundary of Minkowski space.  The loss and
gain of states in the bulk corresponds to boundary conditions imposed
at the null infinities, ${\cal I}^+$ and ${\cal I}^-$, of the RS
brane.

\subsection{AdS$_4$ branes}
\label{sec-adsc}

The causal diamond method is readily applied to the more complicated
metric of the AdS$_4$/AdS$_5$ system.  In this case one finds the
relation
\begin{equation}
\arctan e^c - \arctan e^{r_P} =\frac{\Delta t(r_P)}{2} = 
\arctan e^{\rho(r_P)} - \frac{\pi}{4}.
\label{eq-cdads}
\end{equation}
This implies
\begin{equation}
\rho(r_P) = \ln
\frac{1+e^c-e^{r_P}+e^{c+r_P}}{1-e^c+e^{r_P}+e^{c+r_P}}.
\label{eq-relsize}
\end{equation}
A useful equivalent expression is
\begin{equation}
\rho(r_P) = \ln \frac{\cosh x + \sinh y}{\cosh x - \sinh y},
\end{equation}
where
\begin{equation}
x = \frac{c+r_P}{2},~~~~y=\frac{c-r_P}{2}.
\end{equation}

It is easy to check that $\rho(r_P)=0$ for $r_P=c$,
and
\begin{equation}
{d\rho(r_P) \over d(c-r_P)} >0
\end{equation}
for $r_P<c$.  Thus we find again that the image of a bulk event
at $P$ is a CFT shell whose size increases monotonically with the
distance of the particle from the brane.  

However, the functional relations (\ref{eq-rsrelsize}) and
(\ref{eq-relsize}) differ crucially in that the shell on the AdS$_4$
brane becomes infinitely large already for finite $r_P$.  The function
(\ref{eq-relsize}) diverges at $r_P=b_P$, where
\begin{equation}
\sinh b_P = \frac{-1}{\sinh c}.
\end{equation}
Comparison with (\ref{eq-cau}) reveals that $b_P=b$.

In other words, the CFT shell reaches the boundary of AdS$_4$
precisely when the bulk excitation reaches the caustic surface that
delimits the holographic bulk domain associated to the AdS$_4$ brane!
We have thus completed an important consistency check.  We have
related the escape from the holographic domain to the evanescence of
the holographic image.  The holographic image of a bulk particle
exiting the brane's holographic domain, ${\cal D}(H)$, is a CFT shell
moving off the brane.

\section{Global holography of the AdS$_4$/AdS$_5$ system}
\label{sec-disc}

After a particle leaves ${\cal D}(H)$, it is still in an AdS$_5$ bulk,
and one would expect a kind of holographic duality to a CFT to
persist.  Since this CFT cannot be on the brane, it must reside
elsewhere.  In this section we find that simple bulk dynamics leads to
an interesting interplay between two field theories, or equivalently,
between a single field theory and its boundary.

For definiteness, let us consider a particle released at time $t=0$
from the origin, $\rho=0$, of an AdS$_4$ brane at $r=c$.  The particle
will fall freely in the bulk, which means that it will oscillate
between $r=c$ and $r=-c$.  Its frequency is $\ell^{-1}$ as measured at
$r=0$, or $\ell_4^{-1}$ as measured on the brane at $r=c$.  In
particular, coming from the brane, it will not stop and turn around
when it reaches the boundary of ${\cal D}(H)$, at $r = b$.  (In this
section we will take $c\gg 1$ so that (\ref{eq-cau}) can be
approximated as $b \approx -e^{-2c} \approx 0$.)  Hence, the
holographic image on the AdS$_4$ brane, after moving off to spatial
infinity, will {\em not\/} be reflected back into the brane.
\EPSFIGURE{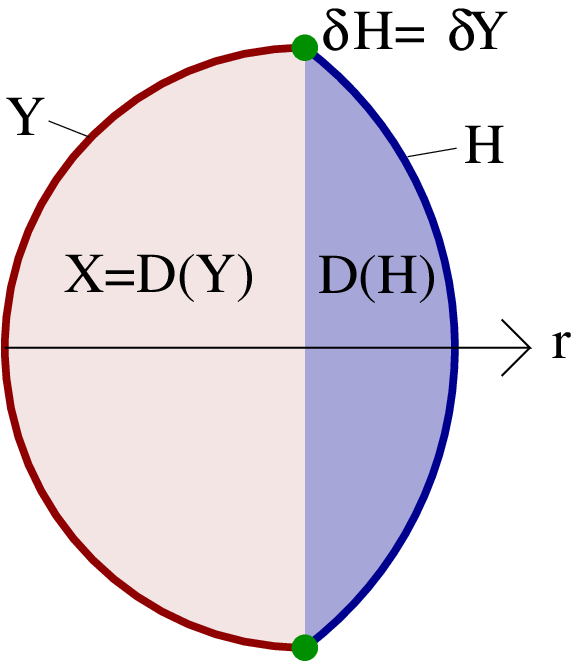,width=6cm}%
{Constant time slice through the Penrose diagram of the
AdS$_4$/AdS$_5$ system.  The $S^2$ angular directions are suppressed
for clarity.  They can be partially recovered by rotating the diagram
around the $r$ axis.  The diagram shows the brane, $H$, its
holographic domain, ${\cal D}(H)$, the surviving portion of the
AdS$_5$ boundary, $Y$, and its holographic domain, ${\cal D}(Y)$.
\label{fig-globslice}}

This is significant, because it is common to impose reflecting
boundary conditions in Anti-de~Sitter space-times.  In the case of the
AdS$_4$ brane, this would mean that the shell should be instantly
reflected at spatial infinity and start to shrink back.  We have just
shown that this behavior would be inconsistent with the bulk dynamics.
We will now show that consistent boundary conditions on the AdS$_4$
brane are obtained by coupling the brane to the remaining boundary of
the AdS$_5$ space.%
\footnote{It was already observed in~\cite{KarRan00} that the
AdS$_4$/AdS$_5$ system has a dual description in terms of two
conformal field theories interacting on their intersection.  This was
also true of the intersecting brane configurations discussed
in~\cite{KarRan01b,DewFre01}.  What is new here is our use of causal
diamonds, which allow us to describe details of the CFT dynamics dual
to certain bulk processes.}

The metric (\ref{eq-adswarp}) covers AdS$_5$ globally, with $r$
ranging from $-\infty$ to $\infty$.  In the AdS$_4$/AdS$_5$ system a
brane resides at $r=c$.  The bulk metric is still described by
(\ref{eq-adswarp}), but $r$ ranges only from $-\infty$ to $c$
(Fig.~\ref{fig-globslice}).  In Sec.~\ref{sec-domains}, we showed that
the portion $0<r<c$ constitutes the holographic domain of the AdS$_4$
brane, ${\cal D}(H)$.  The remaining region, $X$, is defined by
$-\infty < r <0$.  $X$ is half of AdS$_5$, and has a boundary, $Y$, at
$r \rightarrow -\infty$.  $Y$ has topology $D^3 \times {\mathbb R}$,
i.e., its spatial sections are three-dimensional conformal disks.
Note that $Y$ and $H$ are joined on a surface $\partial Y=\partial H$
of topology $S^2 \times R$, defined by taking $\rho \rightarrow
\infty$ at arbitrary fixed $r$.

It is convenient to regard $Y$ as the $c' \rightarrow -\infty$ limit
of an AdS$_4$ hypersurface at $r=c'$.  By substituting $c'$ for $c$ in
(\ref{eq-cau}) and taking the limit, we find that the light-sheets of
$Y$ have caustics at $r=0$.  Hence, the holographic domain of $Y$ is
precisely the region $X$:
\begin{equation}
{\cal D}(Y) = X.
\end{equation}
One would expect that an AdS/CFT duality relates bulk excitations with
support in $X$ to CFT states on $Y$.

We have used the light-sheet construction to show that the entire
AdS$_5$ bulk is the disjoint union of the holographic domain of the
brane, ${\cal D}(H)$, with the holographic domain of the remaining
portion of the boundary, ${\cal D}(Y)$.  Next, we will use causal
diamonds to verify that a particle crossing the boundary between
${\cal D}(H)$ and ${\cal D}(Y)$, $r=0$, corresponds to a CFT
excitation crossing the boundary between $H$ and $Y$, $\partial
H=\partial Y$.

For a particle in ${\cal D}(H)$, we found in the previous section that
its holographic image becomes larger on $H$ and reaches $\partial H$,
as the particle moves away from the brane and reaches $r=0$.  What
remains to be verified is that the image of a particle in ${\cal
D}(Y)$ lies on $Y$ and approaches $\partial Y$ as the particle
approaches $r=0$ from the other side.

Not only the light-sheet construction, but also the discussion of
causal diamonds carries over to hypersurfaces at $r=c'$, with $c'
\rightarrow -\infty$.  Substituting in (\ref{eq-cdads}) we find that a
bulk particle at $r=r_P<0$ will be dual to a shell of radius
\begin{equation}
\rho(r_P) = \ln \coth \frac{|r_P|}{2}
\end{equation}
on $Y$.  When the particle approaches the boundary of ${\cal D}(Y)$,
at $r=0$, we find $\rho \rightarrow \infty$.  Then the shell will be
on $\partial Y$.  This is significant, because the boundary of $Y$ is
also the boundary of the AdS$_4$ brane:
\begin{equation}
\partial Y=\partial H.
\end{equation}
This allows the holographic image to cross over smoothly between $Y$ and
$H$ as a bulk particle crosses $r=0$.

Now let us return to the bulk particle oscillating between $r=\pm c$.
Coming from the AdS$_4$ brane $H$, at $r=c$, it will fall towards
$r=0$.  The holographic dual to this part of its path is a CFT shell
expanding on $H$.  The shell reaches infinity, $\partial H$, when the
particle gets to the boundary of the brane's holographic domain,
${\cal D}(H)$, near $r=0$.  Then the particle crosses over to the bulk
region $X={\cal D}(Y)$, and the shell moves from $\partial Y$ onto $Y$
proper.  The particle turns around at $r=-c$, when the shell is at
$\rho= \ln \coth (c/2)$ on $Y$.

We now understand why reflecting boundary conditions are inappropriate
for the spatial infinity of the AdS$_4$ brane.  The brane $H$ is
coupled at $\partial H$ to a portion of the conformal infinity of
AdS$_5$, $Y$, which can be thought of as the limit of another AdS$_4$
space.  Instead of reflecting at $\partial H$, CFT data first complete
a half-period in $Y$ before re-entering $H$.  It will be an
interesting challenge to understand how such a coupling can be
realized.  

Ultimately, the entire physics on $H$ should be dual to a $2+1$
dimensional CFT on $\partial Y$.  This reduces the holographic dual of
the global bulk space-time to a CFT on $Y$ with boundary dynamics at
$\partial Y$. This would follow from the conjectured duality in
Ref.~\cite{KarRan01b} which has been partially verified in
Ref.~\cite{DewFre01}.  In this picture, the dynamics we have found
would take a particularly striking form.  All physics in ${\cal D}(H)$
would be dual (via $H$) to data on the boundary $\partial Y$ of the
CFT on $Y$.  Excitations on $Y$ typically reach $\partial Y$ and get
stuck there for a time equal to the light-crossing time of $Y$.
Hence, the boundary dynamics at $\partial Y$ must be highly
non-trivial.

In Ref.~\cite{KarRan00} it was noted that four-dimensional gravity
should break down at sufficiently large scales on the AdS$_4$ brane.
This does not happen in the case of the RS brane, where 4D gravity is
valid at all scales.  The difference can be ascribed to the divergence
of the warp factor in the bulk region $X$.  But in Sec.~\ref{sec-adsc}
we showed that an experiment of sufficient duration to probe the
entire brane is just at the threshold to probing $X$.  This suggests
that four-dimensional gravity does not break down at finite {\em
distances\/} on the AdS$_4$ brane.  Rather, the relevant scale is a
time scale.  An experiment longer than $\pi \ell_4$ will detect the
growing warp factor beyond $r=0$, and hence will see a region where
gravity appears genuinely five-dimensional.  It is significant that
$X$ also coincides with the bulk region outside the brane's
holographic domain, as we discuss in the final section.

\section{Interpretation}
\label{sec-interpret}

In both the RS and the KR models, the holographic domain of the full
boundary is obviously the entire bulk space.  Four-dimensional gravity
localizes near the brane, and bulk physics is expected to be dual to
CFT excitations on the brane.  In the KR model, however, the boundary
consists of two distinct portions: the AdS$_4$ brane at $r=c$, and the
half-AdS$_5$ boundary at $r=-\infty$.

In order to interpret the regime that exhibits four-dimensional
gravity, it is necessary to separately identify the holographic domain
associated with each portion of the boundary.  A naive, space-like
relation between brane and bulk data cannot distinguish domains and
would violate the holographic entropy bound.  However, we were able to
obtain distinct domains by using the light-sheet construction central
to the covariant formulation of the holographic principle.  We found
that the domain of the brane extends only to the minimum of the warp
factor, at $r=0$.  The remainder of the bulk, $r<0$, coincides with
the holographic domain of the half-AdS$_5$ boundary.

We developed a technique based on causality that allowed us to locate
the holographic dual on the boundary as a function of the position of
a bulk excitation.  In the RS model, we found that the causal diamond
method reproduces previous results obtained by more intricate means.
In the KR model, we were able to perform an important consistency
check.  The holographic image of a bulk excitation, as constructed via
causal diamonds, respects our association of holographic domains to
the two portions of the boundary.  The image always lies on the
portion of the boundary in whose holographic domain the bulk
excitation is localized.

Generic physical processes, such as freely falling particles, thus
correspond to highly nontrivial dynamics on the boundary.  From the
global perspective, we see that bulk excitations can cross from one
holographic domain to the other, corresponding to conformal field
theory excitations interacting over the ``boundary of the boundary'',
$\partial H$.

An alternative description takes the point of view of the observer on
the brane.  Experiments longer than $\ell$ and shorter than $\ell_4$
will measure four-dimensional gravitational effects.  Experiments of
greater duration are able to probe $X$, the bulk region outside the
brane's holographic domain.  Hence they are able to detect information
that cannot be contained in the CFT on the brane.  The brane observer
can interpret this in two ways.  One perspective is to note that such
experiments probe beyond the minimum of the warp factor and see that
gravity is five-dimensional at large time-scales.  A second option is
to retain the four-dimensional perspective: The observer lives in an
AdS$_4$ space with non-standard boundary conditions, namely not
reflective boundary conditions, but those determined by a CFT on
$Y$. We can think of the AdS$_4$ brane as being glued across its
conformal boundary to (the $\ell_4 \rightarrow \infty$ limit of)
another AdS$_4$ space.  Five-dimensional gravity is re-interpreted as
dynamics in a conformal AdS$_4$ space-time adjacent to the AdS$_4$
brane.  For example, the half-period part of an oscillating bulk
particle's path that lies in $X$ corresponds to its holographic image
dipping into and out of that second (conformal) AdS$_4$ space.

Finally, one may also describe the entire AdS$_4$/AdS$_5$ system as a
CFT on $Y$ with non-trivial boundary dynamics on $\partial Y$.  Such a
theory would have to be able to retain excitations on the boundary for
a long but finite period, corresponding to the time that bulk
excitations spend within ${\cal D}(H)$.

With the discovery and detailed study of novel space-time backgrounds,
we not only witness the emergence of four-dimensional gravity under
ever more general conditions, but we also learn new lessons that might
yield fundamental insights. In this paper, we have seen that the
consistency of holography in the KR model is upheld by the covariant
formulation of the holographic principle, providing, in turn, support
for this formulation and new insights into its workings.

There is much left to be understood about gravity. The discovery of
new geometries has already revealed characteristics previously thought
to be inadmissible.  For example, the
existence~\cite{KarRan00,KogMou00b} and
consistency~\cite{Por00,KogMou00a,KarKat01} of a massive graviton was
not anticipated but is now fairly well understood. In this paper, we
have done a detailed study of another new and surprising aspect of the
AdS$_4$/AdS$_5$ geometry, namely the coexistence of different
holographic domains reflecting different dimensionality. It is now
clear that new geometries can give rise to fascinating new phenomena
which might well provide critical clues to addressing the nature of
gravity.

\acknowledgments

We would like to thank O.~DeWolfe, A.~Karch, L.~Susskind, and R.~Wald
for useful conversations.  This research was supported in part by the
National Science Foundation under Grant No.\ PHY99-07949.


\bibliographystyle{board}
\bibliography{all}

\end{document}